\documentclass[twocolumn,prl]{revtex4}
\usepackage{graphicx}
\usepackage{amssymb,amsmath,bm}
\usepackage{color}

\newcommand{\be}{\begin{equation}}
\newcommand{\ee}{\end{equation}}
\newcommand{\bea}{\begin{eqnarray}}
\newcommand{\eea}{\end{eqnarray}}

\newcommand{\la}{\langle}
\newcommand{\ra}{\rangle}

\newcommand{\lp}{\left(}
\newcommand{\rp}{\right)}

\renewcommand{\phi}{\varphi}
\renewcommand{\epsilon}{\varepsilon}

\begin{document}

\title{Coherent Particle Transfer in an On-Demand Single-Electron Source}

\author{J.~Keeling$^1$, A.~V.~Shytov$^2$, L.~S.~Levitov${}^{3,4}$}
\affiliation{
$^1$ Cavendish Laboratory, University of Cambridge,
 Madingley Road, Cambridge CB3 0HE, UK \\
$^2$ Department of Physics, University of Utah, Salt Lake City, UT 84112 \\
${}^3$ Kavli Institute for Theoretical Physics,
University of California Santa Barbara, CA 93106 \\
${}^4$ Department of Physics, Massachusetts Institute of
  Technology, 77 Massachusetts Ave, Cambridge, MA 02139}

\begin{abstract}
Coherent electron transfer from a localized state trapped in a quantum
dot into a ballistic conductor, taking place in on-demand electron
sources, in general may result in excitation of particle-hole pairs.
We consider a simple model for these effects, involving a resonance
level with time-dependent energy, and derive Floquet scattering matrix
describing inelastic transitions of particles in the Fermi sea.  We
find that, as the resonance level is driven through the Fermi level,
particle transfer may take place completely without particle-hole
excitations for certain driving protocols. In particular, such
noiseless transfer occurs when the level moves with constant rapidity,
its energy changing linearly with time. A detection scheme for
studying the coherence of particle transfer is proposed.
\end{abstract}
\maketitle

Individual quantum states of light, supplied on demand by
single-photon sources \cite{imamoglu94,brunel99}, are essential for
current progress in manipulating and processing quantum information in
quantum optics \cite{qubits}. In particular, such sources are at the
heart of secure transmission of quantum information by quantum
cryptography \cite{bennett92}, and of quantum teleportation
\cite{Bouwmeester97}.  An extension of these techniques to electron
systems would be crucial for the inception of fermion-based quantum
information processing \cite{beenakker03,samuelsson04}.

While a number of elements of solid state electron optics, such as
linear beamsplitters \cite{wees88,wharam88} and interferometers
\cite{ji03}, have been known for some time, an on-demand electron
source was demonstrated only recently \cite{glattli07}. In the
experiment \cite{glattli07} a localized state in a quantum dot,
tunnel-coupled to a ballistic conductor, was controllably charged and
discharged by time-dependent modulation of the energy of the state
induced by a periodic sequence of voltage pulses on the gate.  In this
process electrons are alternatingly, one at a time, injected in
(trapped from) a quantum Hall edge channel, leading to a sequence of
quantized single-electron current pulses \cite{Moskalets2008}.  The
energy of the injected electron could be independently controlled by
tuning the out-coupling of the dot.

Yet, the nearly perfect quantization of current pulses achieved in
\cite{glattli07} in general does not guarantee full quantum coherence.
In a fully coherent pulse, the injected electron occupies a prescribed
quantum state without accompanying particle/hole pairs excited from
the Fermi sea. However, since particle/hole pairs have a finite
density of states at low energy, a generic perturbation applied to a
Fermi system is expected to create multiple pairs.  This process,
which has no analog for photon sources, constrains the protocols for
generating coherent pulses.

\begin{figure}% [htbp]
  \centering
  \includegraphics[width=0.62\columnwidth]{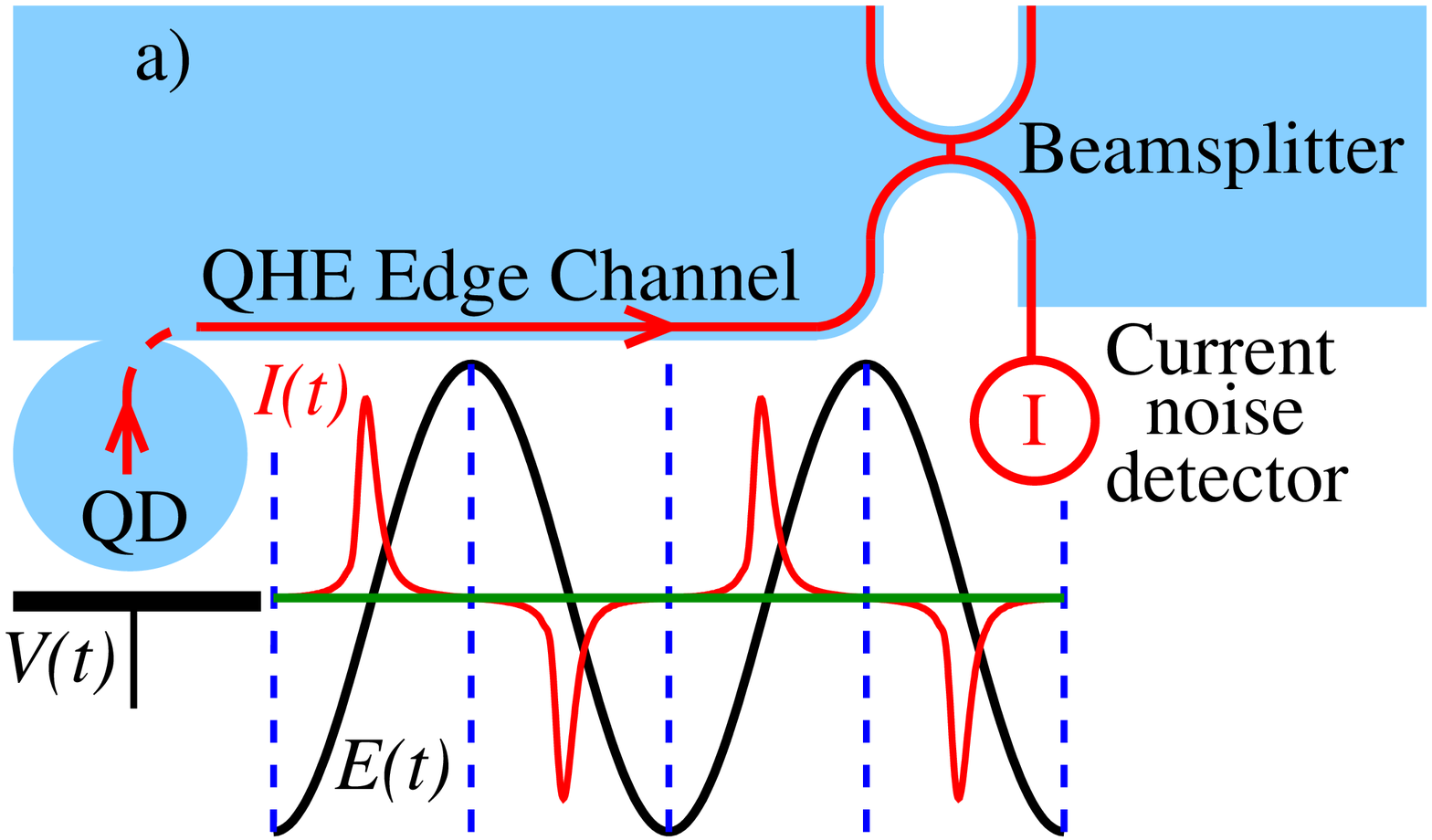}
\includegraphics[width=0.35\columnwidth]{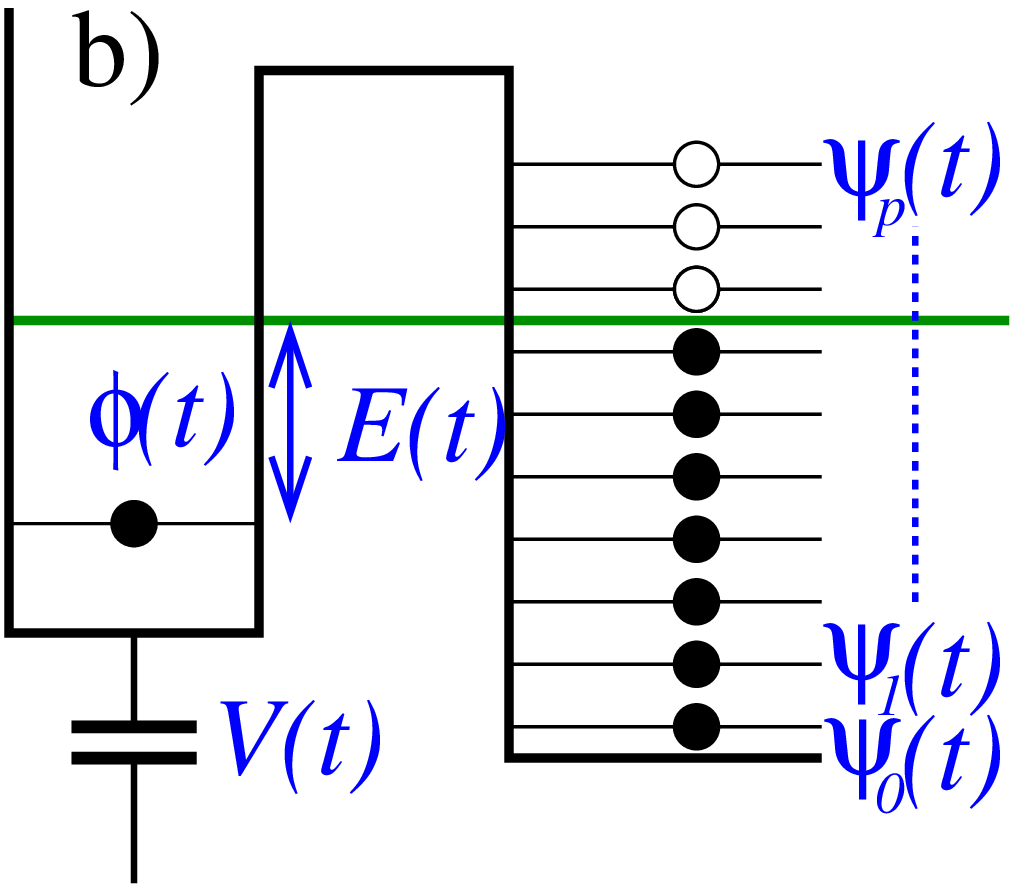}
\caption{a) Quantum dot tunnel-coupled to a ballistic
  conductor. Electrons are periodically trapped on the dot and
  injected in the conductor as the electron energy $E(t)$ in the dot
  is increased above the Fermi level when a time-dependent voltage
  $V(t)$ is applied to the gate. Particle-hole excitations which
  accompany the injected electron can be detected by the current
  partition noise on a beamsplitter.  b) Schematic diagram of a
  localized level coupled to a continuum of propagating modes,
  Eq.\eqref{eq:H}.  }
  \label{fig:general}
\end{figure}

To characterize the coherence of particle transfer, we employ an exact
time-dependent (Floquet) scattering matrix, generalizing the
Breit-Wigner theory of resonance scattering to arbitrary time
dependence of the localized state energy $E(t)$.  Applying this
approach to the many-body evolution of a Fermi sea coupled to a
localized state with driven energy, we identify the case of linear
driving $E(t)=ct$, in which excitation creation is fully
inhibited. The harmonic driving used in \cite{glattli07} is well
approximated by this linear model if electron release and capture
occur well within each half-period, as shown in
Fig.\,\ref{fig:general}\,a.  For such clean protocols the entanglement
between injected particle and the Fermi sea is totally suppressed by
Pauli blocking of multi-particle excitations.

Clean protocols are not restricted to the adiabatic limit, and so one
may study the form of clean current profiles as a function of speed of
driving, which interpolates between Lorentzian when adiabatic, and
exponential when fast, with fringes for intermediate rates.  We also
study how robust such protocols are to imperfections expected in
experiment, such as noise in the driving voltage.  Our results can
also be relevant for quantum pumps (see
\cite{Geerligs90,Blumenthal07,Moskalets02} and references therein).

A method to distinguish optimal and non-optimal protocols is
illustrated in Fig.~\ref{fig:general}\,a, by measuring the shot noise
from current partitioning on a beamsplitter downstream of the electron
source.  For non-optimal protocols, the total number of excitations
$N^{\mathrm{ex}}$ (electrons + holes) is greater than one.  Because
every electronic state scatters independently, the variance of
transfered charge depends on the number of excitations that may
scatter.  Hence, the DC shot noise generated on the beamsplitter is
$e^2 {\sf t}(1-{\sf t})N^{\mathrm{ex}}\nu$, where ${\sf t}$ is the
beamsplitter transmission coefficient and $\nu$ is the frequency of
current pulses.

In the setup of Ref.\cite{glattli07} the gate used to vary $E(t)$ is
placed so close to the dot that the charging energy $e^2/2C$ is small
compared to level spacing, which allows to leave out the Hubbard-like
interaction term.  Also, because magnetic field of a few Tesla was
applied to create a Quantum Hall state in which electron spins are
polarized, only one spin projection is considered, hence electron
transfer from a quantum dot to the Fermi sea is described by the
many-body Hamiltonian (Fig.\ref{fig:general}\,b):
\be\label{eq:H}
{\cal H} = E(t)d^\dagger d + \sum_p \epsilon_p a^\dagger_p a_p + \lambda_p(t)d^\dagger a_p + \lambda_p^\ast(t)a_p^\dagger d
\ee
where $d$ and $a_p$ describe the localized and extended states. 
Here $E(t)$ is the time dependent electron energy in the dot, and $\lambda(t)$ is the tunneling amplitude, for generality also taken to be time dependent.

To describe time evolution of (\ref{eq:H}) we shall first find the
single-particle scattering matrix for transitions among the continuum
states $|p\ra$. For that, we must solve the Schr\"odinger equations
for the propagating modes $\psi_p(t)$ coupled to the wavefunction
$\phi(t)$ of the localized state:
\[
\left[i \partial_t - \epsilon_p \right] \psi_p = \lambda_p^{\ast}(t) \phi
,\quad
\left[i \partial_t - E(t) \right] \phi = \sum_p \lambda_p(t) \psi_p
\]
(we set $\hbar=1$ and $\epsilon_F=0$ unless specified otherwise).
Crucially, because the localized state is coupled to the continuum at
all times, its behavior (e.g. charging or discharging) can be fully
accounted for by an S-matrix for transitions in the continuum.  The
situation here is completely analogous to the Breit-Wigner theory of
resonance scattering in which an energy-dependent scattering phase is
used to describe the resonance.

Because the continuum of propagating QHE modes in \cite{glattli07} is
one-dimensional, it is convenient to go over to position
representation $\psi(t,x) = \sum_p e^{ipx}\psi_p(t)$. Hereafter we
assume a constant density of states and treat the couplings
$\lambda_p$ as energy independent.  Replacing $\epsilon_p$ by $ - i
v_{F}\partial_x$, where $v_F$ is the Fermi velocity, gives
\bea
  \label{eq:eom-phi}
  && \left[ i \partial_{t} - E(t) \right] \phi(t) 
  = 
  \lambda(t) \int dx \delta(x) \psi(t,x)
  \\
  \label{eq:eom-psi}
  && \left[ i \partial_{t} + i v_F \partial_x \right] \psi(t,x) 
  =
  \lambda^{\ast}(t) \delta(x) \phi(t)
.
\eea
The scattering matrix for energy-nonconserving time evolution can be
labeled by pairs of energies of the continuum states, as
$U(\epsilon,\epsilon^{\prime})$.  Because the continuum modes
propagate freely at $x<0$, the initial state is:
\begin{displaymath}
  \psi(t,x<0)=\psi_0(t,x)
=  e^{- i \epsilon^{\prime}\tilde t}
,\quad \tilde t=t-\frac{x}{v_F},
\end{displaymath}
and $\phi(t=-\infty)=0$ as discussed above. Projecting the evolved
state onto an equivalent final state gives
\begin{equation}
  \label{eq:definen-U}
  U(\epsilon,\epsilon^{\prime}) = 
  \int  dt
  \psi(t,x>0)  
e^{i \epsilon\tilde t}
.
\end{equation}
Let us now solve the coupled equations of motion, and thus
find $U(\epsilon,\epsilon^{\prime})$.
First solving Eq.~(\ref{eq:eom-psi}), one finds:
\begin{equation}
  \label{eq:soln-psi}
  \psi(t,x) = \psi_0(\tilde t)-
  \frac{i}{v_F} 
  \lambda^{\ast}(\tilde t)
  \phi(\tilde t)
  \theta(x).
\end{equation}
Substituting this into Eq.~(\ref{eq:eom-phi}) 
we find an equation for the localized state:
\begin{equation}
  \label{eq:eom2-phi}
  \left[ i \partial_t - E(t) + i \frac{\gamma(t)}{2}\right]
  \phi(t) = 
  \lambda(t)  \psi_0(t)
,
\end{equation}
where we introduced notation $\gamma(t)=|\lambda(t)|^2/v_F$ for the
localized level linewidth.
Then, the solution of Eq.~(\ref{eq:eom2-phi}) with the initial
condition $\phi(-\infty)=0$ is of the form
\begin{equation}
  \label{eq:soln-phi}
  \phi(t) = -i  \int_{-\infty}^{t} dt^{\prime} 
  \lambda(t^{\prime}) \psi_0(t^{\prime})
  e^{X(t,t')}
,
\end{equation}
where $X(t,t')=-\int^t_{t'}\lp \frac12\gamma(\tau)+iE(\tau)\rp d\tau$.
The result \eqref{eq:soln-phi} may be substituted into
Eq.\eqref{eq:soln-psi} for $\psi(t,x)$; this can in turn be used in
Eq.~(\ref{eq:definen-U}) to evaluate $U(\epsilon,\epsilon^{\prime})$.
Putting all this together, we find the Floquet S-matrix
\bea
  \label{eq:general-U}
&&  U(\epsilon,\epsilon^{\prime}) = \iint dt dt' e^{i\epsilon t-i\epsilon' t'}U(t,t'),
\\\nonumber
&& U(t,t')=\delta(t-t')-\theta(t-t')\frac{\lambda^\ast(t)\lambda(t')}{v_F}  
e^{X(t,t')}
.
\eea
It is straightforward to show that $U$ is unitary, $U^\dagger U=\hat 1$, by verifying that $\int U(\tau,t)U^\ast(\tau,t')d\tau = \delta(t-t')$. 

As a sanity check, let us apply these results to a stationary
level. For time-independent $\gamma$ and $E$, we find $X(t,t')=-\lp
\frac{\gamma}2+iE\rp (t-t')$. Integrating over $t$ and $t'$ in
\eqref{eq:general-U}, we obtain the familiar result:
\be\label{eq:Breit-Wigner}
U(\epsilon,\epsilon')=2\pi\delta(\epsilon-\epsilon') 
\frac{\epsilon-E-i\gamma/2}{\epsilon-E+i\gamma/2}
.
\ee
A more interesting example is a level moving at a constant rapidity,
$E(t)=ct$. In this case, $X(t,t')=-\frac{\gamma}2(t-t')-\frac{ic}2
(t^2-{t'}^2)$. After integrating over $t$ and $t'$ in
\eqref{eq:general-U} we find $U(\epsilon,\epsilon') = 2\pi
\delta(\epsilon-\epsilon')+ T(\epsilon,\epsilon')$ where
\be
  \label{eq:linear-U}
  T(\epsilon,\epsilon')=-2\pi\frac{\gamma}{|c|}\theta\lp \epsilon-\epsilon'\rp 
e^{
  - \frac{\gamma}{2c}(\epsilon - \epsilon')
  + \frac{i}{2c}(\epsilon^2 - \epsilon^{\prime 2})
  }
\ee
for $c>0$, and with $\theta\lp \epsilon'-\epsilon\rp$ instead of $\theta\lp \epsilon-\epsilon'\rp$ for $c<0$.
This result agrees with the continuum limit of the Demkov-Osherov S-matrix \cite{demkov68,Sinitsyn2002} for a single level crossing a group of stationary levels.

We next employ this single-particle scattering matrix in the calculation
of the many body properties, taking as the initial state the filled Fermi
sea.
The number of excitations can be obtained from the initial filled
Fermi sea state $|\Omega\rangle$ evolved with the S-matrix
$U(\epsilon,\epsilon^{\prime})$.  In particular, the number of
fermions promoted above the Fermi level is:
\begin{equation}
  \label{eq:nfermions}
  N^{+} =
  \left<\Omega\right|U^{\dagger} 
  \!\sum_{\epsilon>0}\! a_{\epsilon}^{\dagger} a_{\epsilon}
  U \left| \Omega\right>
  =
\sum\limits_{\epsilon>0,\, \epsilon'<0}  \left| U(\epsilon,\epsilon^{\prime})\right|^2
.
\end{equation}
Similarly, the number of holes created below the Fermi surface $N^{-}$
is found by swapping $\epsilon$ and $\epsilon^{\prime}$ in Eq.\eqref{eq:nfermions}.

Using our explicit expression for $U(\epsilon,\epsilon^{\prime})$ (but
assuming now $\gamma(t) = \gamma$ for simplicity), one may rewrite the
result \eqref{eq:nfermions} by using $\int_0^\infty d\epsilon
e^{i\epsilon(t-s)}= \frac{i}{t-s+i 0}$, which yields:
\be
  \label{eq:nfermi-et}
  N^{+} =
-\lp\frac{\gamma}{2\pi}\rp^2\iiiint\limits_{t>t',\,s>s'} 
  \frac{e^{X(t,t')+X^\ast(s,s')}dt dt' ds ds'
  }{%
    (t-s + i 0)(t^{\prime} - s^{\prime} + i 0)
  }
\ee
(replacing $i0$ by $-i0$ gives an expression for $N^-$).  It can be
seen from these expressions that in general the numbers of excited
particles and holes are not constrained.

To illustrate this, let us first consider a highly non-optimal
protocol for E(t), where the level first moves rapidly to the
Fermi-level, remains there for time $\Delta t$, and then moves rapidly
away.  During the time $\Delta t$, the level acts as a resonant
perturbation for the Fermi sea, with scattering phase
$\delta(\epsilon)=\tan^{-1}\frac{2(\epsilon-E)}{\gamma}-\frac{\pi}2$
defined by \eqref{eq:Breit-Wigner} This creates a logarithmically
divergent number of excitations, $N^{\pm} \propto \log \Delta t$,
which can be understood as an example of the ``orthogonality
catastrophe'' \cite{Anderson67}.

The situation is completely different in the case when the level moves
linearly, $E(t)=ct$. From our result for the S-matrix,
Eq.\eqref{eq:linear-U}, we have $U(\epsilon<\epsilon')=0$ for $c>0$.
This means that no holes are excited when the level is moving up in
energy: $N^-=0$ for $c>0$.  (Similarly when the level is moving down,
i.e. $c<0$, one has $U(\epsilon>\epsilon')=0$ and thus $N^+=0$).  At
the same time, we expect that $N^{+}-N^{-}=+1(-1)$ when the level
moves up (down) as just one particle is transfered between the
localized level and continuum. Indeed, we can find $N^{\pm}$ directly,
by substituting $U$ into Eq.~(\ref{eq:nfermions}) yielding $N^{+}=1$
for $c>0$ (and $N^{-}=1$ for $c<0$).  Thus for linear driving, a
single fermion is coherently transfered from the localized level to
the continuum, and no holes are created.

\begin{figure}  % [htbp]
  \centering
  \includegraphics[width=0.98\columnwidth]{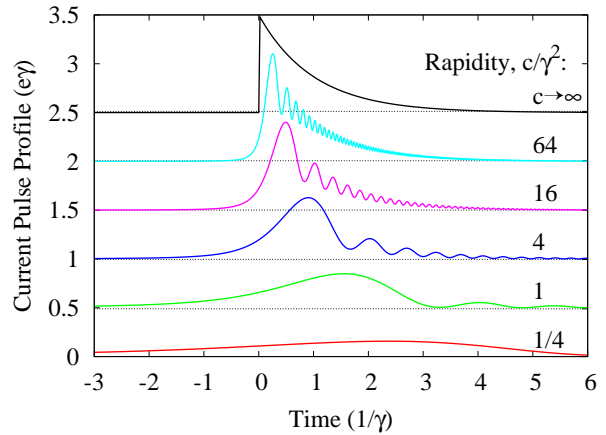}
  \caption{Single electron current pulse produced by linear driving of a localized state of width $\gamma$ across the Fermi level, $E(t)=ct$. The pulse profile $|\psi(\tilde t)|^2$, $\tilde t=t-x/v_F$, Eq.\eqref{eq:linear-state}, is shown for different values of rapidity $c$.
Different curves are offset by $0.5 e\gamma$. The asymptotic form is Lorentzian at slow driving, and exponential at fast driving. Note interference fringes at the trailing side of the pulse at $c\gtrsim\gamma^2$.
} 
  \label{fig:pulse_profile}
\end{figure}

This remarkable behavior can also be understood directly from
Eq.~\eqref{eq:linear-U}: restricted to $\epsilon>0$, $\epsilon'<0$, $T(\epsilon,\epsilon')$ is a rank one matrix.
As discussed in Ref.~\cite{keeling06}, for a rank one S-matrix the exact many body state is a product of an unperturbed Fermi sea and an extra particle occupying one mode, which is a superposition of harmonics with $\epsilon>0$. The latter can be read off directly from \eqref{eq:linear-U}:
\begin{equation}
  \label{eq:linear-state}
  \psi(t,x)= \sqrt{\frac{\gamma}{2\pi c}}
  \int_{0}^\infty \!\! d\epsilon \, \exp\left[
    -i \epsilon \tilde t
    - \frac{\gamma\epsilon}{2c} + i \frac{\epsilon^2}{2c}\right].
\end{equation}
This gives a density profile $|\psi(x,t)|^2$ that is the convolution
of a Lorentzian, width $\gamma/c$, with a Fresnel integral, leading to
fringes on the trailing side of the pulse (Fig.\ref{fig:pulse_profile}).

The key features of particle transfer under linear driving will be
similar to that under periodic driving, used in Ref.\cite{glattli07},
if the latter sweeps a wide enough interval of energies on either side
of the Fermi level.  In particular, if the period is long compared to
$\max (\gamma^{-1},\gamma/c)$, and the extremal value of $E(t)$
exceeds $\gamma$, as indicated in Fig.\ref{fig:general}a; then
particle transfer will be nearly noiseless, close to that under linear
driving.  In addition, comparison to Ref.\cite{demkov68} indicates
that particle transfer will remain noiseless in a more general case,
when the tunnel coupling and the density of states are energy
dependent, as in \eqref{eq:H}.

\begin{figure}  % [htbp]
  \centering 
\includegraphics[width=0.9\columnwidth]{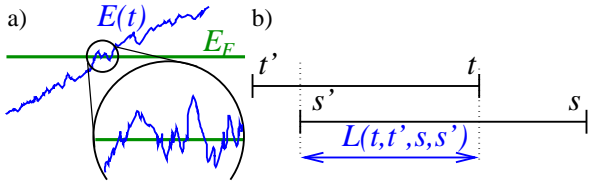}
\caption{ a) Multiple crossing of the Fermi level in the presence of
  noise, Eq.\eqref{eq:define-noise}, leads to creation of
  particle/hole excitations; b) Overlap of time intervals and
  definition of $L(t,t^{\prime},s,s^{\prime})$.  }
  \label{fig3}
\end{figure}

More insight into the robustness of the coherent particle transfer can
be gained by considering, as an example, the effect of classical noise
added to $E(t)$ (for a discussion of Landau-Zener transitions in the
presence of different kinds of noise see
\cite{Kayanuma98,Pokrovsky03,Wubs2006} and references therein).
In experimental realizations, the energy of the localized level is not
under perfect control;  as well as the desired applied voltage, there
will be Johnson noise and noise associated with fluctuating charges.
For simplicity we consider the effect of noise on the linear driving
case (see Fig.~\ref{fig3}a):
\begin{equation}
  \label{eq:define-noise}
  E(t) = ct + \xi(t),
\end{equation}
where $\left<\xi(t)\right>=0$ and
$\left<\xi(t)\xi(t^{\prime})\right>=\gamma_2 \delta( t-t^{\prime})$.
Substituting this into the equation for number of excitations,
Eq.~(\ref{eq:nfermi-et}), and averaging over realizations of the
noise, we find that the integrand can be written as:
\begin{equation}
  \label{eq:integrand}
    \frac{%
    e^{
      - \frac{\gamma}{2}(t-t^{\prime} + s - s^{\prime})
      -
      \frac{i c}{2} (t^2 - t^{\prime 2} - s^2 + s^{\prime 2} )
    }
  }{%
    (t-s + i 0)(t^{\prime} - s^{\prime} + i 0)
  }
  \times
  F(t,t^{\prime},s,s^{\prime})
,
\end{equation}
where the factor $F = \la e^{i \int_{s^{\prime}}^{s} \xi(\tau) d\tau - i \int_{t^{\prime}}^{t} \xi(\tau) d\tau} \ra$ equals
\be\label{eq:noiseavg}
\exp\lp -\frac{\gamma_2}{2} \left[
      |t-t^\prime|
      +
      |s-s^\prime|
      - 2 L(t,t',s,s^{\prime})
    \right]\rp
.
\ee
Here $L(t,t^{\prime},s,s^{\prime})$ is the overlap between the two
intervals $[t,t^{\prime}]$ and $[s,s^{\prime}]$ 
(see Fig.~\ref{fig3}b).
This simple form \eqref{eq:noiseavg}
comes from the Gaussian correlations of
$\xi(\tau)$ leading to cancellation for any region inside $L(t,t^{\prime},s,s^{\prime})$.

To make further progress, we change variables to
$s=t+\eta$, $t^{\prime}=t-\Delta_t$, 
$s^{\prime}=s-\Delta_s$.
Because $F$ does not depend on central time $t$,
the only $t$ dependence in (\ref{eq:integrand}) comes from 
$
t^2 - t^{\prime 2} - s^2 + s^{\prime 2} 
  =
  \Delta_t (2t - \Delta_t) - 
  \Delta_s(2t+2\eta - \Delta_s).
$
Integration over $t$ thus gives a delta function, $\frac{2\pi}{c}
\delta(\Delta_t - \Delta_s)$, which considerably 
simplifies the
expression:
\begin{equation}
  \label{eq:simplified-Nex}
    N^{+}
    = 
    \frac{-\gamma^2}{2\pi c}
    \int_{-\infty}^{\infty} \!\!\! d \eta
    \int_0^{\infty} \!\!\! d {u}
    \frac{e^{ (-ic\eta - \gamma){u} + \gamma_2 (L(\eta, {u})-{u})}}{
      (\eta+i0)^2},
\end{equation}
(where we have written $\Delta_t=\Delta_s={u}$).
In terms of $u$ and $\eta$,
the overlap can be written in the simpler
form:
\begin{math}
  L(\eta,{u}) = \theta({u} - |\eta|)({u} - |\eta|)=u-\mathrm{min}({u}, |\eta|).
\end{math}
We can now examine the asymptotic limits of expression
\eqref{eq:simplified-Nex}; for this it is convenient to consider the
total number of excitations $N^{\mathrm{ex}} = N^+ + N^-$.
[Due to overall conservation of fermions, and assuming an initially
populated localized state, $N^+ - N^-=1$, as discussed above.]
Combining the form of $N^{+}$ in Eq.~(\ref{eq:simplified-Nex}) and the
matching form for $N^-$ with $i0\to-i0$, one may use the identity
$(\eta+i0)^{-2}+(\eta-i0)^{-2}=-\int_{-\infty}^{\infty} \!\!\! d\omega |\omega| e^{i\eta\omega}$.
After relabeling $\omega$ to $\omega+ c {u}$, we obtain
\be\nonumber
  N^{\mathrm{ex}} = 
  \frac{\gamma^2}{2\pi}
  \int\limits_{-\infty}^{\infty} \!\! d \omega
  \int\limits_{-\infty}^{\infty} \!\! d \eta
  \int\limits_{0}^{\infty}  \!\! d {u}
\,\textstyle{  \left| {u} + \frac{\omega}{c} \right|}\,
e^{i\omega\eta - \gamma {u} - \gamma_2 \,\mathrm{min}({u},|\eta|)}
.
\ee
In this form, it is easy to extract the asymptotic limits of
fast and slow driving.
At large $c$, we may approximate $|{u} + \omega/c|
\approx {u}$.
Then, integration over $\omega$ yields a delta function,
$2\pi\delta(\eta)$, giving $ \lim_{c\to \infty} N^{\mathrm{ex}} = 1$.
This limit has a simple interpretation; if driven fast enough, the
effects of noise do not matter, and one recovers the clean case
discussed earlier.

In the limit of small $c$, we retain only the terms proportional
to $1/c$.
Defining $\gamma_{\ast}=\gamma +\gamma_2$, we may write:
\begin{equation}
  \label{eq:small-c-int-lambda}
  \int_{0} ^{\infty}  \!\! d {u}
  e^{%
    - \gamma {u} - \gamma_2\, \mathrm{min}({u},|\eta|)
  }
 =
  \frac{1}{\gamma_{\ast}} + \frac{\gamma_2}{\gamma \gamma_{\ast}} 
  e^{-\gamma^{\ast} |\eta|}
\end{equation}
The integration over $\eta$ then gives
\be
  \label{eq:}
  N^{\mathrm{ex}} =
  \frac{2\gamma\gamma_2}{\pi c}
  \int_{0}^{\infty}\!\!\! d\omega \frac{\omega}{\omega^2 + \gamma_{\ast}^2}
  \approx
  \frac{2\gamma\gamma_2}{\pi c}
  \ln \frac{\omega_0}{\gamma_{\ast}} 
.
\ee
In the final expression, we have introduced a high $\omega$ cutoff
$\omega_0$, to remove the ultraviolet divergence; this divergence
corresponds to short time correlations.
The origin of this divergence is the white noise spectrum for
$\xi(t)$; the divergence relates to the fact that for a truly white
spectrum, there will be an infinite number of crossings of the Fermi
level, and so an infinite number of excitations.
By comparing the fast and slow driving limits, we find the crossover
occurs at the rapidity $c = (2/\pi) \gamma\gamma_2 \ln(\omega_0/\gamma_{\ast})$.

In conclusion, excitation of particle/hole pairs in a single-electron
source can be suppressed by optimizing the protocol of particle
transfer between a localized state and continuum. The transfer is
totally noiseless when the energy of the localized state varies
linearly in time. In this case, owing to the Fermi statistics,
particle/hole pair production is suppressed by Pauli blocking of
multi-particle excitations. The quantum state resulting from such
clean transfer is a product state of a particle added to an
unperturbed Fermi sea, with zero entanglement between them.
Particle/hole excitation, and its suppression, can be observed
directly by noise measurement.

\begin{acknowledgments}
%%  We would like to thank 
We are grateful to Christian Glattli and Israel Klich for useful discussion.
 J.K. acknowledges financial support from the Lindemann Trust,
 and Pembroke College Cambridge. L.L.'s work was partially supported by W. M. Keck foundation and by the NSF grant PHY05-51164.
\end{acknowledgments}


\begin{thebibliography}{99}


\bibitem{imamoglu94}
A. Imamoglu, Y. Yamamoto, 
%% Turnstile device for heralded single photons...
Phys. Rev. Lett. {\bf 72}, 210 (1994).

\bibitem{brunel99}
C. Brunel, B. Lounis, Ph. Tamarat, M. Orrit,
%% Triggered Source of Single Photons based on Controlled Single Molecule Fluorescence
Phys. Rev. Lett. {\bf 83}, 2722 (1999).

\bibitem{qubits}
P. Kok, {\it et al.},
%% W. J. Munro, K. Nemoto, T. C. Ralph, J. P. Dowling, and G. J. Milburn, 
Rev. Mod. Phys. {\bf 79}, 135 (2007).

\bibitem{bennett92}
C. H. Bennett, {\it et al.}, 
%% F. Bessette, G. Brassard, L. Salvail, and J. Smolin, 
%% "Experimental quantum cryptography," 
J. Cryptology {\bf 5}, 3 (1992).

\bibitem{Bouwmeester97}
D. Bouwmeester, {\it et al.}, 
%% J. Pan, K. Mattle, M. Eibl, H. Weinfurter, and A. Zeilinger, 
%% "Experimental quantum teleportation," 
Nature {\bf 390}, 575 (1997).


\bibitem{beenakker03}
C. W. J. Beenakker, C. Emary, M. Kindermann, J. L. van Velsen, Phys. Rev. Lett. {\bf 91}, 147901 (2003).

\bibitem{samuelsson04}
P. Samuelsson, E. V. Sukhorukov, M. B\"uttiker, Phys. Rev. Lett. {\bf 92}, 026805 (2004).

\bibitem{wees88}
%Quantized conductance of point contacts in a two-dimensional electron gas 
B. J. van Wees, {\it et al.}, 
%% H. van Houten, C. W. J. Beenakker, J. G. Williamson, L. P. Kouwenhoven, D. van der Marel, C. T. Foxon,
Phys. Rev. Lett. {\bf 60}, 848 (1988).

\bibitem{wharam88}
D. A. Wharam, {\it et al.}, 
%% T. J. Thornton, R. Newbury, M. Pepper, H. Ahmed, J. E. F. Frost, D. G. Hasko, D. C. Peacockt, D. A. Ritchie, G. A. C. Jones,
%% D. A. Wharam et al., 
%% One-dimensional transport and the quantisation of the ballistic resistance
J. Phys. C {\bf 21}, L209 (1988).

\bibitem{ji03}
Y. Ji, {\it et al.},
%% Y. Chung, D. Sprinzak, M. Heiblum, D. Mahalu, H. Shtrikman,
%% An electronic Mach\u2013Zehnder interferometer
Nature {\bf 422}, 415 (2003).

\bibitem{glattli07}
G. F\`{e}ve, {\it et al.}, 
%% A. Mah\'{e}, J.-M. Berroir, T. Kontos, B Placais, D. C. Glattli, A. Cavanna, B. Etienne, Y. Jin,
%% An On-Demand Coherent Single Electron Source
Science {\bf 316}, 1169 (2007). 

\bibitem{Moskalets2008}
M. Moskalets, P. Samuelsson, M. Buttiker, 
%% Quantized dynamics of a coherent capacitor
Phys. Rev. Lett. {\bf 100}, 086601 (2008).

\bibitem{Geerligs90} 
L. J. Geerligs, {\it et al.}, 
%% V. F. Anderegg, P. A. M. Holweg, J. E. Mooij, H. Pothier, D. Esteve, C. Urbina, and M. H. Devoret, 
%% Frequency-locked turnstile device for single electrons
Phys. Rev. Lett. {\bf 64}, 2691 (1990).

\bibitem{Blumenthal07}
M. D. Blumenthal, {\it et al.}, 
%% B. Kaestner, L. Li, S. Giblin, T. J. B. M. Janssen, M. Pepper, D. Anderson, G. Jones, D. A. Ritchie,
%% Gigahertz quantized charge pumping
Nat. Phys. {\bf 3}, 343 (2007) 

\bibitem{Moskalets02}
M. Moskalets and M. B\"uttiker,
%% Dissipation and noise in adiabatic quantum pumps
Phys. Rev. B{\bf  66}, 035306 (2002);
%% \bibitem{Moskalets02}
%% M. Moskalets and M. B\"uttiker,
%% Floquet scattering theory of quantum pumps
Phys. Rev. B{\bf  66}, 205320 (2002).


\bibitem{demkov68}
Y. N. Demkov, V. I. Osherov, 
%% Stationary and nonstationary problems in quantum mechanics that can be solved by contour integration
Zh. Eksp. Teor. Fiz {\bf 53}, 1589 (1967) [Eng. translation: Sov. Phys. JETP {\bf 26}, 916 (1968)].

\bibitem{Sinitsyn2002}
N. A. Sinitsyn,
Phys. Rev. B{\bf 66}, 205303 (2002).
%% Multiparticle Landau-Zener problem:\u2003Application to quantum dots

\bibitem{Anderson67} P. W. Anderson, Phys. Rev. Lett. {\bf 18}, 1049 (1967).

\bibitem{keeling06} 
J. Keeling, I. Klich, L. S. Levitov, Phys. Rev. Lett. {\bf 97}, 116403 (2006).

%% \bibitem{Saito02}
%% K. Saito, Y. Kayanuma, Phys. Rev. A{\bf 65}, 033407 (2002).

\bibitem{Kayanuma98}
Y. Kayanuma, H. Nakayama,
%% Nonadiabatic transition at a level crossing with dissipation
Phys. Rev. B{\bf 57}, 13099  (1998).

\bibitem{Pokrovsky03}
V. L. Pokrovsky, N. A. Sinitsyn, Phys. Rev. B{\bf 67}, 144303 (2003). 

\bibitem{Wubs2006}
M. Wubs, K. Saito, S. Kohler, P. H\"anggi, and Y. Kayanuma,
%% Gauging a Quantum Heat Bath with Dissipative Landau-Zener Transitions
Phys. Rev. Lett. {\bf 97}, 200404 (2006).

\end{thebibliography}
\end{document}